\documentclass{emulateapj}
\RequirePackage{lineno}
\usepackage{amssymb}
\usepackage{amsmath}
\usepackage{mathrsfs}
\usepackage{natbib}
\usepackage{array}
\usepackage{float}
\usepackage{graphicx}
\usepackage{subfigure}
\usepackage{color}

\def\beq{\begin{equation}}
\def\enq{\end{equation}}
\def\bea{\begin{eqnarray}}
\def\ena{\end{eqnarray}}

\def\jcap{Jour. Cosmology and Astro-Particle Phys.\,}

\shorttitle{Supernova and Hypernova Contributions to High Energy Backgrounds}
\shortauthors{Xiao et al.}

\begin{document}

\title{Revisiting the Contributions of Supernova and Hypernova Remnants to the Diffuse High-Energy Backgrounds:
Constraints on Very-High-Redshift Injection}
\author{Di Xiao\altaffilmark{1,2,3}, Peter M\'esz\'aros\altaffilmark{3}, Kohta Murase\altaffilmark{3} and Zi-Gao Dai\altaffilmark{1,2}}
\affil{\altaffilmark{1}School of Astronomy and Space Science, Nanjing University, Nanjing 210093, China}
\affil{\altaffilmark{2}Key Laboratory of Modern Astronomy and Astrophysics (Nanjing University), Ministry of Education, China}
\affil{\altaffilmark{3}Center for Particle and Gravitational Astrophysics; Department of Physics; Department of Astronomy and Astrophysics, The Pennsylvania State University, University Park, PA 16802, USA}

\date{\today}

\begin{abstract}
Star-forming and starburst galaxies are considered as one of the viable candidate sources of
the high-energy cosmic neutrino background detected in IceCube.  We revisit contributions of supernova
remnants (SNRs) and hypernova remnants (HNRs) in such galaxies to the diffuse high-energy neutrino
and gamma-ray backgrounds, in light of the latest {\it Fermi} data above 50~GeV.  We also take into
account possible time dependent effects of the cosmic-ray (CR) acceleration during the SNR evolution.
CRs accelerated by the SNR shocks can produce high-energy neutrinos up to $\sim100$ TeV energies,
but CRs from HNRs can extend the spectrum up to PeV energies.
We show that, only if HNRs are dominant over SNRs, the diffuse neutrino background above 100 TeV can
be explained without contradicting the gamma-ray data. However, the neutrino data around 30 TeV remain
unexplained, which might suggest a different population of gamma-ray dark CR sources.
We also consider possible contributions of Pop-III HNRs up to $z\lesssim10$, and show that
they are not constrained by the gamma-ray data, and thus could contribute to the diffuse high-energy backgrounds
if their explosion energy reaches ${\mathcal E}_{\rm POP-III}\sim{\rm a~few}\times10^{53}$~erg.
More conservatively, our results suggest that the explosion energy of POP-III HNRs is
${\mathcal E}_{\rm POP-III}\lesssim7\times{10}^{53}$~erg.
\end{abstract}


\section{Introduction}
During last few years the IceCube experiment has detected tens of high-energy astrophysical neutrinos
from 10 ${\rm TeV}$ to a few ${\rm PeV}$ \citep{aar13a,aar13b,aar15}. This outstanding observation result has led
to a wide debate about the origin of these neutrinos, \citep[e.g.,][]{mes14,mur14nu}.  Among various source models,
the most commonly discussed are SNRs (like W49B \citep{lop13, gon14}) in star-forming galaxies (SFGs) and starbursts galaxies (SBGs)
\citep[e.g.,][]{loe06, mur13, liu14, tam14, anc14, cha14, cha15, nick15}, which are good cosmic-ray (CR) reservoirs and
can produce neutrinos via $pp$ interactions.
At the same time, this interaction will produce a certain amount of gamma-rays and contribute to the extragalactic
gamma-ray background (EGB), and a common origin of the diffuse backgrounds has been suggested \citep{mur13}.
Other possible astrophysical sources which can contribute to the IceCube neutrino flux include
gamma-ray bursts (GRBs) \citep{wax97, mw01, mur08ph, wan09, bus14, xia14, xia15, ta16},
low-luminosity GRBs or trans-relativistic supernovae (SNe)~\citep{mur06, gz07, mura13, senno16},
active galactic nuclei \citep{ste91, alv04, anc08, ste13, mur14, der14, pet15}, galaxy clusters and groups
\citep{mur08}, newborn pulsars \citep{mur09, fan14}.  Alternatively, there are also arguments about a
possible Galactic contribution \citep[e.g.,][]{fox13, am14, tay14, ner16}.

\par
On the other hand, the Fermi-LAT instrument has measured the EGB in the energy range from 0.1 to 820
GeV \citep{ack15}.  New studies of the blazar flux distribution at gamma-ray energies above 50 GeV have
shown that most of the EGB flux should be contributed by blazars, at a percentage level $86_{-14}^{+16}\%$
\citep{bec15,dim16}. This means that the diffuse gamma-ray flux from other sources cannot exceed the
residual non-blazar component, which puts a strong constraint on the SBG scenario \citep{mur13}.
\citet{mur16} found that SBGs cannot explain the IceCube data below 100 TeV especially if the above
constraint on the non-blazar component is taken into account.
More quantitatively, \citet{bec15} argued, from the known branching ratio between charged and neutral
pion production, that the gamma-ray emission of SBGs explaining the IceCube data would be well beyond
$14\%$, so these observational results would be in strong tension with the SBG scenario.
Interestingly, similar results are also found by the gamma-ray anisotropy data \citep{and16}.

In this work, we argue that if we attribute, as is common, the production of CRs in SBGs to SNRs,
then taking into account the time-behavior of the SNR evolution, and taking into account also a
contribution from very-high-redshift SNRs in a Pop-III component, it is possible to reduce the portion
of the EGB which is in tension with Fermi, so that SBGs could contribute a potentially interesting fraction
of the IceCube neutrino flux.

We assume that the neutrinos are mainly produced at the end of the free expansion and during the
Sedov-Taylor phase of the SNR, in which the maximum CR energy and maximum neutrino energy naturally
decreases with time due to a decreasing expansion velocity and magnetic field in the shock. For the
accompanying $\gamma$-rays, after the $\gamma\gamma$ cascade process, this also translates into a
different ``elbow" of the EGB.

The total CR energy injection rate during the evolution may be uncertain, and here we parametrize
it as a simple universal power law in time during the Sedov phase, $L_{\rm CR} \propto t^\alpha$,
with $\alpha$ ranging from zero to negative values. We then discuss the impact of $\alpha$ on the
final neutrino and gamma-ray spectrum. We found that negative values of $\alpha$ are more favorable
from the point of view of attributing a significant fraction of the neutrino and gamma-ray
background to SNRs, including both normal HNe and SNe, as discussed below. Conversely, an
appreciable contribution of SBG SNRs to these backgrounds would have implications for the acceptable
values of $\alpha$, and thus for the CR production efficiency as a function of the SNR dynamics.

One main difference between our paper and previous work is that we allow for a wider range of
HN ejecta energies, and we allow for the different CR maximum energy and flux contribution
at different stages of the SNR expansion.
The other, more significant major difference is that we allow for a Pop-III population of SNRs,
with a comparable cosmic ray acceleration efficiency and comparable or larger kinetic energy
as those attributed to $z\lesssim 4$ SNRs and HNRs.

This paper is organized as follows. We introduce the model and method of calculation in Section 2.
Then section 3 presents our results showing the combined fits of our model compared with the IceCube
and Fermi EGB data. The implications of the results are presented in Section 4.

\section{Model and Method of Calculation}
\subsection{Sedov-Taylor Phase of Supernova Remnant}

After the early free expansion, the evolution of a SNR enters the Sedov-Taylor self-similar expansion
phase \citep{sed59,tay50}. The ejected remnant shell collides with and sweeps up the ambient medium,
forming a shock which can accelerate CRs that subsequently, via $pp$ collisions, produce high energy
neutrinos and $\gamma$-rays. We take the beginning of the Sedov-Taylor phase as the onset time
for CR acceleration and secondary neutrino and $\gamma$-ray production. This occurs at a time
\bea
t_0=\frac{R_{\rm dec}}{v_{\rm {ej}}}&=&\frac{(3M_{\rm ej}/4\pi m_pn_0)^{1/3}}{(2\mathcal{E}_{\rm SN}/M_{\rm ej})^{1/2}}\nonumber\\
&\simeq&1.4\times10^3\mathcal{E}_{\rm SN,51}^{-1/2}n_{0,0}^{-1/3}M_{\rm ej,1}^{5/6}\,{\rm yr}
\label{eq:t0}
\ena
after the initial explosion. The ending time of the Sedov-Taylor phase $t_{\rm end}$ occurs when the
adiabatic expansion approximation is no longer true, and this is determined by equating the cooling time
of the post-shock region with the SNR dynamic time. The gas cooling function
has been evaluated in detail by \citep[e.g.][]{sut93}, and has been adopted in our calculation
to solve for $t_{\rm end}$ as a function of different sets of parameters.
\par
We take the usual time dependence of the radius and the shock velocity in the Sedov-Taylor phase
$R=({25\mathcal{E}_{\rm SN}}/{4\pi m_pn_0})^{1/5}t^{2/5}$ and
$v=({2}/{5})({25\mathcal{E}_{\rm SN}}/{4\pi m_pn_0})^{1/5}t^{-3/5}$ \citep{sed59, tay50, ram10},
and the commonly used estimate for the post-shock magnetic field $B=(9\pi\epsilon_Bm_pn_0v^2)^{1/2}$
where $\epsilon_B \leq 1$ is the ratio of the post-shock magnetic to thermal energy.
In this work, we will assume the CR component to be pure protons. Then, for diffusive shock acceleration,
the maximum proton energy is estimated by equating the local acceleration time to the
dynamical time (the cooling timescale for $pp$ collisions and synchrotron losses being much longer
than the dynamical timescale, e.g. \cite{nick15}), which gives $({20\epsilon_{p,\rm max}c}/{3eB})=Rv$
\citep{dru11}. Thus, in the Sedov-Taylor phase the maximum CR energy also depends on time,
\bea
\epsilon_{p,\rm max}=1.58\times10^6\mathcal{E}_{\rm {SN},51}n_{0,0}^{1/6}M_{\rm ej,1}^{-2/3}\epsilon_{B,-2}^{1/2}(t/t_0)^{-4/5}~{\rm GeV} \,\,\,\,\,
\label{eq:Epmax}
\ena

\subsection{Energy injection into CRs}

In the free expansion phase, for an external density $\rho=$ constant, as long as the swept-up
mass is small the velocity $v\simeq$ constant and the radius grows as $R\propto t$. The rate
at which mass is swept and is shock-heated is ${\dot M}\propto R^2 \rho v \propto R^2\propto t^2$,
and the bolometric luminosity that can go into CRs grows as $L\propto {\dot M} \propto t^2$.
The equipartition post-shock field behaves as $B\propto v\propto$ constant, and the shock strength
is also constant. Thus, at the simplest level one expects the CR bolometric luminosity to grow as
$L_{\rm CR}\propto t^2$.  In the Sedov phase, on the other hand, $R\propto t^{2/5}$ and $v\propto
t^{-3/5}$ so one expects ${\dot M}\propto t^{1/5}$, while $B\propto v \propto t^{-3/5}$, and with
decreasing velocity the shock strength decreases. Thus, one does not expect a positive proportionality
between ${\dot M}$ and $L_{\rm CR}$ but instead a decrease of $L_{\rm CR}$ can be expected.
The SN ejecta begins to slow down at $t_0$ when it has swept up an amount of external mass
comparable to that of the ejecta shell, marking the beginning of the Sedov-Taylor phase, when the
cosmic ray acceleration is expected to peak.
The details of the continued CR injection as the blast wave slows down in the Sedov-Taylor phase
are model-dependent, and here we shall parametrize it as
\beq
L_{\rm CR}(t)=\mathcal{A}(t/t_0)^{\alpha},
\label{eq:alpha}
\enq
where $\mathcal{A}$ is a normalization factor determined by
$\int_{t_0}^{t_{\rm end}} L_{\rm CR}dt=\eta \mathcal{E}_{\rm SN}$.
Here $\eta\sim0.1$ is the CR acceleration efficiency \citep{cs15}.
According to equation (\ref{eq:alpha}), at different stages of the Sedov-Taylor phase, the energy
injection rate into CRs will be different.  The value $\alpha=-1$ is a nominal steady injection
case \citep{ohi10}. Because of the very fast growth $L_{\rm CR} \propto t^2$ in the
initial  free expansion phase, for simplicity we neglect contributions from this phase.
We shall assume the shock-accelerated proton spectrum to be a power law
$dN_p/d\epsilon_p\propto\epsilon_p^{-p}$ with index $p=2$, so at time $t$, the proton spectrum
being written as:
\beq
   \epsilon_pL_{\epsilon_p}(t)=
   \begin{cases}
   \frac{L_{\rm CR}(t)}{\mathcal{C}} &\mbox{if $\epsilon_p\leq \epsilon_{p,\rm max}(t)$}\\
   \frac{L_{\rm CR}(t)}{\mathcal{C}}e^{-\epsilon_p/ \epsilon_{p,\rm max}} &\mbox{if $\epsilon_p> \epsilon_{p,\rm max}(t)$}
\end{cases}
\label{eq:spectrum}
\enq
where $\mathcal{C}=\ln{(\epsilon_{p,\rm max}/\epsilon_{p,\rm min})}$

\subsection{Diffuse Neutrino and Gamma-ray Fluxes}

Hypernovae (HNe) are a sub-type of supernovae (SNe) which are more energetic but relatively rarer.
Recent studies indicate that the total rate of all core-collapse supernovae (CCSNe) is $\mathcal{R}_{\rm CCSNe}=
(1.06\pm0.11(stat)\pm0.15(sys))\times10^{-4}(h_0/0.7)^3 {\rm Mpc^{-3}yr^{-1}}$ \citep{tayo14}.
For typical HNe, the kinetic energy is of order $10^{52-53}{\rm erg}$, with a typical rate $\mathcal{R}_{\rm HNe}
\leq 4\%\mathcal{R}_{\rm CCSNe}$ \citep{gue07, arc10, nick15}, but with substantial uncertainties. HNe have been discussed as possible sources for high-energy cosmic rays \citep{der02, sve03, wan07, iok10}.
Due to larger kinetic energy, HNe can lead to larger maximum proton energies and will dominate the neutrino flux at PeV energies.
\par
The total local CR energy budget is
\bea
\epsilon_pQ_{\epsilon_p}(z=0)&=&\mathcal{R}_{\rm SNe}\int_{t_{\rm 0,SNe}}^{t_{\rm end,SNe}}\epsilon_pL^{\rm SNe}_{\epsilon_p}(t)dt\nonumber\\
&+&\mathcal{R}_{\rm HNe}\int_{t_{\rm 0,HNe}}^{t_{\rm end,HNe}}\epsilon_pL^{\rm HNe}_{\epsilon_p}(t)dt
\label{eq:Q0}
\ena
and the cosmological evolution of sources can be expressed as
$\epsilon_pQ_{\epsilon_p}(z)=\epsilon_pQ_{\epsilon_p}(z=0)S(z)$, in which the evolution term is
$S(z)=[(1+z)^{-34}+(\frac{1+z}{5000})^3+(\frac{1+z}{9})^{35}]^{-0.1}$ \citep{hop06,yuk08}.
In this work, the typical neutrino energy is approximated to be $E_\nu\sim0.05\epsilon_p/(1+z)$.
\par
The diffuse neutrino flux per flavor is calculated as \citep[e.g.,][]{mur13,nick15,bec15,mur16}
\bea
E_{\nu}^2\Phi_{\nu_i}&=&\frac{c}{4\pi H_0}\int_0^{z_{\rm max}}\frac{\mathrm{min}[1,f_{pp}]}{6}\\\nonumber
&\times&\frac{\epsilon_pQ_{\epsilon_p}(z=0)S(z)}{(1+z)^2\sqrt{\Omega_M(1+z)^3+\Omega_\Lambda}}dz
\label{eq:nuflux}
\ena
where $z_{\rm max}=4$, the cosmology parameters are $H_0=67.8\,{\rm km\,s^{-1}\,Mpc^{-1}}, \Omega_M=0.308$
\citep{pla15} and
$f_{pp}=\bar{n}\kappa\sigma_{pp}c\cdot \mathrm{min}[t_{\rm diff},t_{\rm adv}]$. Here the diffusive escape
timescale of CRs from the host galaxy is $t_{\rm diff}=h^2/4D$, where $h$ is the galaxy scale-height,
the diffusion coefficient is $D(\epsilon_p)=D_c[(\epsilon_p/\epsilon_{p,c})^{1/3}+
(\epsilon_p/\epsilon_{p,c})^2]$ and the timescale of advective escape is $t_{\rm adv}=h/V_w$. We use
the values of $h=10^{21}\rm {cm}$, $D_c\sim3.1\times10^{29}\rm {cm^2\,s^{-1}}$,
$\epsilon_{p,c}\sim9.3\times10^9\rm {GeV}$, $V_w=1500\rm km\,s^{-1}$, $B=4\rm mG$,
$\bar{n}=100\rm cm^{-3}$ for SBGs \citep{str09} and $h=1\rm {kpc}$, $D_c
\sim3.1\times10^{29}\rm {cm^2\,s^{-1}}$, $\epsilon_{p,c}\sim9.3\times10^6\rm {GeV}$,
$V_w=500\rm km\,s^{-1}$, $B=1\rm\mu G$, $\bar{n}=1\rm cm^{-3}$ for SFGs \citep{kee06, cro12}.
We assume a starburst fraction of $\xi_{\rm SBG}=0.1$ \citep{tam14}. Note that the upstream magnetic
field of SBGs can be stronger than in the usual ISM due to the superbubble structure. Also, the upstream
field amplification by CRs may be expected \citep{ber09}.
A further contribution to $f_{pp}$ from diffusion in the host galaxy cluster is considered in \citet{nick15}.
\par
Based on the standard branching ratio between charged and neutral pion production of $pp$ interaction,
the gamma-ray flux can initially be related to the neutrino flux through
$E_{\gamma}^2\Phi_{\gamma}=2E_{\nu}^2\Phi_{\nu}|_{\epsilon_\nu=0.5\epsilon_\gamma}$.
However, once they are produced, high energy gamma-rays undergo $\gamma\gamma$ interactions. The main target
soft photons are from the extragalactic background light (EBL) \citep[e.g.,][]{ss98,dai02a,dai02b,mur12}.
The optical depth for gamma-rays depends on their energy $\epsilon_\gamma$, as well as on redshift,
because the EBL intensity varies at different $z$ \citep{fin10}. We can set $\tau_{\gamma\gamma}(E_\gamma,z)=1$
to get the cut-off energy $E_\gamma^{\rm cut}$. Beyond $E_\gamma^{\rm cut}$, electromagnetic cascade develops,
and the resulting spectrum has a universal form \citep{bre75,mur12},
\beq
   E_\gamma\frac{dN_\gamma}{dE_\gamma}\propto
   \begin{cases}
   (\frac{E_\gamma}{E_\gamma^{\rm br}})^{-1/2} &\mbox{if $E_\gamma\leq E_\gamma^{\rm br}$}\\
   (\frac{E_\gamma}{E_\gamma^{\rm br}})^{-1} &\mbox{if $E_\gamma^{\rm br}<E_\gamma\leq E_\gamma^{\rm cut}$}
\end{cases}
\label{eq:gammaspec}
\enq
where $E_\gamma^{\rm br}=0.0085(1+z)^2(\frac{E_\gamma^{\rm cut}}{100\rm GeV})^2$  \citep{nick15}.
The normalization is determined by integrating the gamma-ray flux over $E_\gamma^{\rm cut}$.
\par
Below $E_\gamma^{\rm cut}$, the gamma-ray intensity is attenuated by a factor $e^{-\tau_{\gamma\gamma}}$.
This is calculated similarly to the diffuse neutrino flux, with a typical energy $E_\gamma\sim0.1\epsilon_p/(1+z)$,
\bea
E_{\gamma}^2\Phi_{\gamma}&=&\frac{c}{4\pi H_0}\int_0^{z_{\rm max}}\frac{\mathrm{min}[1,f_{pp}]}{3}\\\nonumber
&\times&\frac{\epsilon_pQ_{\epsilon_p}(z=0)S(z)}{(1+z)^2\sqrt{\Omega_M(1+z)^3+\Omega_\Lambda}}e^{-\tau_{\gamma\gamma}(E_\gamma,z)}dz
\label{eq:gamflux}
\ena
The final diffuse gamma-ray flux is the sum of these two components.

\section{Results}

\subsection{The $z\lesssim 4$ contribution}

\begin{figure}
\includegraphics[width=\linewidth]{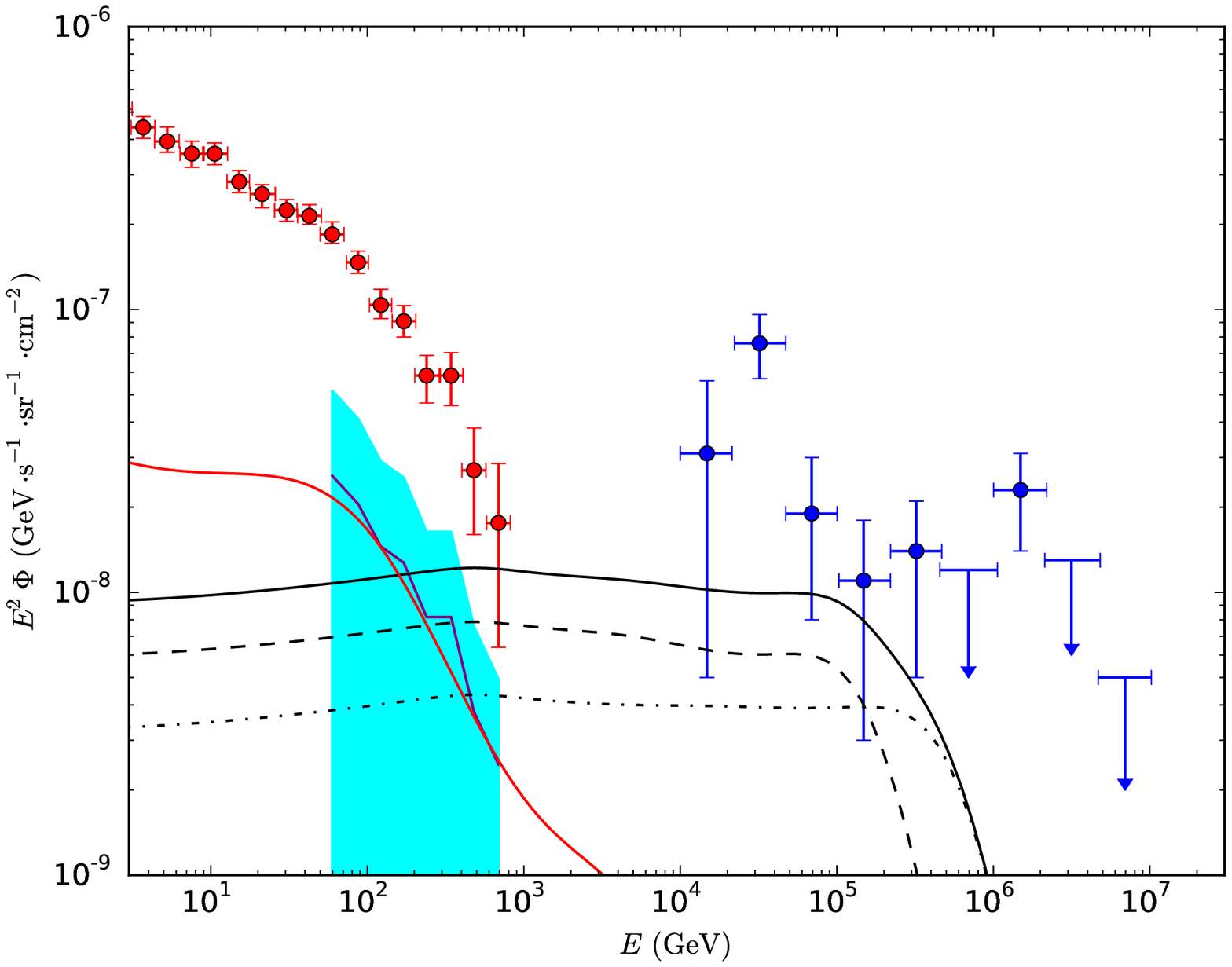}
\caption{Combined fit of diffuse neutrino flux and gamma-ray flux for the case $\alpha=-1$ for the conventional case.
The IceCube neutrino and the Fermi-LAT extragalactic gamma-ray background observations are shown by
blue and red data points respectively \citep{ack15,aar15}.
The cyan area shows the allowed region for the non-blazar gamma-ray flux in \citet{dim16} and the best-fit 14\%
residual of the Fermi EGB is marked by the purple solid line.
Black dashed and dotted lines represent the calculated contribution to the neutrino flux from SNe and HNe
respectively, from the range $z\leq 4$.
The black solid line is the predicted total diffuse neutrino flux and the red solid line is the predicted gamma-ray flux.
The main parameters are $\mathcal{E}_{\rm SNe}=5\times10^{50}$erg, $\mathcal{E}_{\rm HNe}=10^{52}$erg, $\eta=0.1$,
$n_0=1{\rm cm}^{-3}$,$\mathcal{R}_{\rm HNe}=3\%\mathcal{R}_{\rm CCSNe}$.
The SBG magnetic field is set to $B=1$~mG.
\label{fig1}}
\end{figure}

\begin{figure}
\includegraphics[width=\linewidth]{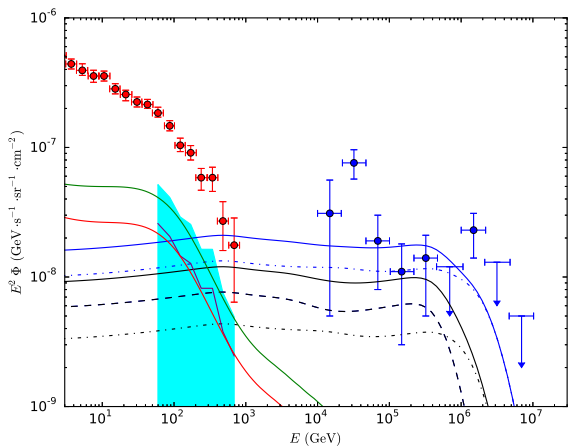}
\caption{The same as Figure~\ref{fig1} but with $\mathcal{E}_{\rm HNe}=10^{52.5}\rm erg$.
The blue solid line is the total neutrino flux, which is the sum of SNe (blue dashed line) and HNe
(blue dotted line) contributions. The green solid line is the diffuse gamma-ray flux in this case.
The SBG magnetic field is set to $B=4$~mG. Black lines show the neutrino spectrum for a
$\mathcal{E}_{\rm HNe}=10^{52}\rm erg$ case for comparison (SNe-black dashed, HNe-black dotted, total-black solid),
while the red solid line is the corresponding diffuse gamma-ray flux.
\label{fig2}}
\end{figure}

Here we use the above discussed values of the luminosity functions, the EBL density and other parameters
which are commonly used for redshifts $z\lesssim 4$.
Figure~\ref{fig1} shows a combined fit of the diffuse gamma-ray and neutrino flux for the nominal case of
$\alpha=-1$ in eq. (\ref{eq:alpha}). We use this nominal value to delimit the physically expected range
of $\alpha \lesssim 0.5$.
The IceCube neutrino and Fermi-LAT extragalactic gamma-ray background (EGB) observation are shown by the
blue and red data points respectively.
The black solid line is the predicted total diffuse neutrino flux and the red solid line represents the gamma-ray flux.
The black dashed and dotted line represent the contribution to the neutrino flux from SNe and HNe respectively.
We can see that HNe dominate the PeV neutrino flux, while SNe only contribute up to sub-PeV neutrino energies.
However, this case based on the conventional assumptions cannot explain the IceCube data, suggesting that the HN rate
would need to be enhanced to explain the neutrino data.
This difficulty in getting a good combined $\nu,\gamma$ fit for this conventional case is compatible with the
independent generic arguments of \citet{bec15} against SBGs being the dominant sources of IceCube neutrinos.
\par
Obviously, we would expect to increase the fraction of HNe contribution because they can achieve higher CR energies.
Figure~\ref{fig2} shows what happens if we increase the HNe kinetic energy to $10^{52.5}$erg. The blue solid line is
the diffuse neutrino flux, which satisfies the observations better now, although it is still below the 1$\sigma$
range of the last data point. The SN contribution is the blue dashed line and the HN are the blue dotted line.
Except for $\mathcal{E}_{\rm HNe}$, all other parameters are the same for the black lines in this figure. The corresponding
gamma-ray flux is the green solid line, which is still in the allowed range. In this case, the ratio between
the HNe and SNe contribution  is $\zeta\equiv\frac{\eta {\mathcal E}_{\rm HN}\mathcal{R}_{\rm HN}}{\eta {\mathcal E}_{\rm SN}\mathcal{R}_{\rm SN}}=\frac{10^{52.5}\times3\%}{5\times10^{50}\times97\%}\sim2$, compared to $\zeta\sim0.6$ for
the $\mathcal{E}_{\rm HN}=10^{52}$erg case. We can draw a rough conclusion that $\zeta\gtrsim1$ is needed for a good fit.
\par
\begin{figure}
\plotone{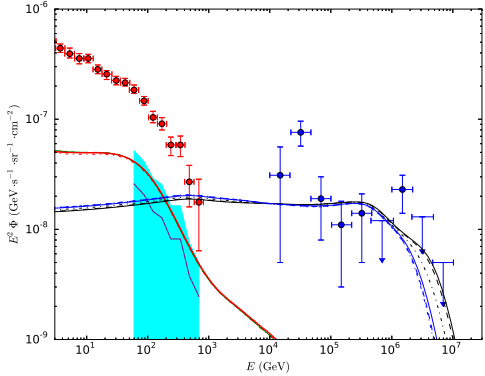}
\caption{Combined fits of diffuse neutrino flux and gamma-ray flux for $z\leq 4$ and different
values of the SNR evolution parameter $\alpha$ of equation (\ref{eq:alpha}).
The IceCube and Fermi-LAT EGB observations are shown by blue and red data points respectively.
All parameters are the same as in Figure~\ref{fig2}, with predicted neutrino fluxes and gamma-ray
fluxes shown by different sets of lines for different $\alpha$.
The values for $\alpha=-3$ are the black solid line and red solid line;
$\alpha=-2$ are the black dashed line and red dashed line;
$\alpha=-1$ are black dotted line and red dotted line;
$\alpha=0$ are blue solid line and green solid line.
For completeness, we also show cases for positive $\alpha$:
$\alpha=1$ are blue dashed line and green dashed line;
$\alpha=2$ are blue dotted line and green dotted line.
\label{fig3}}
\end{figure}

In addition, we consider more realistic cases where $L_{\rm CR} \propto t^\alpha$, for various $\alpha$.
We plot the combined fits for different values of $\alpha$ in Figure~\ref{fig3}. The neutrino spectrum shape
is seen to depend on $\alpha$. Looking at the results for $\alpha=0$ and for positive values $\alpha=1, 2$
(although the positive values are unlikely to occur) we see that all three neutrino spectral curves decrease
sharply at the high energy end, undershooting the data.  One way to address this might be to increase the rate
of HNe (or to choose a very large SBG fraction $\xi_{\rm SBG}$), thus increasing the energy input of HNe so as
to reach the IceCube data, but this inevitably gives rise to an increased gamma-ray flux which would overshoot
the Fermi data.
However, for the physically more reasonable negative $\alpha$ values, the neutrino spectrum
is ``flatter" (since at later times in the HN evolution both the maximum energy and the flux decrease)
and more energy is evident in the flux at the high energy end, so we do not need such a high fraction of
HNe as before. We can see that the last IceCube data point is reachable for $\alpha=-2$ and $\alpha=-3$ using
a quite conservative HN fraction $\mathcal{R}_{\rm HNe}=3\%\mathcal{R}_{\rm CCSNe}$.
Clearly, negative values of $\alpha$ are more favorable from an observational point of view.
\par
\begin{figure}
\plotone{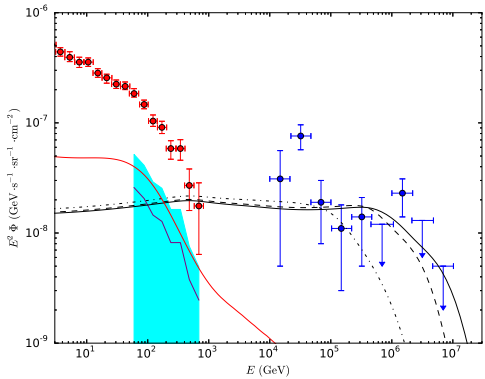}
\caption{The dependence of neutrino flux on the SBG upstream magnetic field, for $z\leq 4$.
We choose the $\alpha=-1$ case as an example.
The dashed line is for $B=4\rm mG$, while the
solid line is for $B=10\rm mG$ (optimistic) and the dotted line is for $B=1\rm mG$.
\label{fig4}}
\end{figure}

Generally, the parameter space gets larger for smaller $\alpha$. The major parameters that we can adjust
in our fits include the SBG upstream magnetic field; the kinetic energies $\mathcal{E}_{\rm SN},
~\mathcal{E}_{\rm HN}$; the pre-shock medium number density $n_0$; the HNe fraction
$\mathcal{R}_{\rm HN}$; and the CR acceleration efficiency $\eta$.  The nominal values used here are
in the observationally reasonable range, e.g. $B=4\rm mG$ \citep[e.g.][]{tho06, rob08, bec16},
$\mathcal{E}_{\rm SN}\sim 10^{50-51}\rm erg, \mathcal{E}_{\rm HN}
\sim 10^{52-53}\rm erg$ \citep[e.g.][]{iwa98, maz03},
$n_0\sim1~\rm cm^{-3}$ \citep{dra91, che01, koo07, fox11}, $\mathcal{R}_{\rm HNe}\leq 0.04\mathcal{R}_{\rm CCSNe}$
\citep{gue07, arc10} and $\eta<1$ (physically required).
For illustration, Figure~\ref{fig4} shows what happens if we change the SBG upstream magnetic field.
A stronger field helps to explain the observations. The key issue here is that we wish to achieve
higher maximum proton energies in order to account for PeV neutrino flux.
%

\par
Alternatively, given the contribution of other sources such as blazars to the Fermi gamma-ray background
\citep{bec15, dim16}, we can provide limits on the SNR contributions to the neutrino flux.
If we normalize the gamma-ray flux to the best-fit 14\% of the total EGB, we find
that for our nominal parameters the corresponding diffuse neutrino flux is about 50\% of the observed
value, provided that we choose not to satisfy the $\sim$ 30 TeV neutrino data point, which however
has small error bars. If this data point has a different physical origin than the higher energy
data points, SBGs may still contribute a potentially interesting fraction of the $\gtrsim 100$ TeV
IceCube neutrino flux.

One may wonder whether very-high-energy gamma-rays may be attenuated inside SFGs and SBGs.
This could happen above $\sim10$~TeV energies \citep{lt10,mur13,cha14}.
However, this would not change the results essentially. The synchrotron losses of the cascade electron-positron
pairs leads to a reduction in the diffuse gamma-ray flux. However, gamma-rays in the 0.1-10 TeV range remain
unattenuated in the sources and will be injected into intergalactic cascades. Also, electron-positron pairs
including the contribution from muon decays as well as primary electrons produce some gamma-rays, which can
make the constraints tighter \citep{lac13}.
%

%

\subsection{A possible Pop-III contribution at $4 \lesssim z \lesssim 10$}

An interesting but more uncertain contribution to the diffuse neutrino and gamma-ray background can
be expected from the Pop-III star formation occurring at high redshifts \citep[e.g.][]{ume03, iwa05}.
For these epochs, besides the increased redshift of the received photons, also (other things being
equal) the ${(1+z)}^3$ increase of the proper densities affects the $\gamma\gamma$ interaction and the
gamma-ray attenuation.
Pop-III stars are extremely metal-poor and can be very massive \citep[e.g.][]{tom02, hos13, wha13, yan15}.
They may end their life as HN explosions, $\mathcal{E}_{\rm kin}=10^{51-53}\rm erg$ \citep[e.g.][]{che14, tom14, tom16},
and such Pop-III SNe, including jet-driven SNe and GRBs, are also expected to accelerate CRs and produce
neutrinos \citep{sch02, iocco07, gao11, ber12}.

\begin{figure}
\plotone{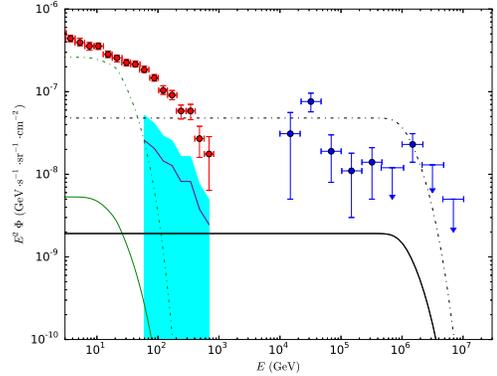}
\caption{Possible contribution from Pop-III HNRs with $4\leq z \leq 10$ by themselves. Black and green
lines represent the predicted diffuse neutrino and gamma-ray fluxes respectively. For the solid lines
the efficiency $\eta$ is 0.1 and the kinetic energy is $\mathcal{E}_{\rm kin}=10^{52.5}$ erg. For the
dotted-dashed lines, the value of $\eta \mathcal{E}_{\rm kin}$ (i.e. the flux) is multiplied by a factor 25.
These dotted-dashed lines serve as an upper bound for the Pop-III contribution.
\label{fig5}}
\end{figure}

The theoretical uncertainties for this Pop-III epoch are significant, but we can use assume reasonable
HN parameters used in the literature for illustrative purposes. For instance, the Pop-III star formation
rate is under debate \citep{wis05, bro06, tre09, des11, wis12} and can be $10-100$ times smaller \citep{tre09, wis12}. 
Here, we use the HN explosion parameters of \citet{che14, tom14} and the EBL model and the upper Pop-III model given in \citet{ino13},
with a typical HN kinetic energy of $\mathcal{E}_{\rm kin}=10^{52.5}$ erg and an external gas density
$n_0=10^3$ cm$^{-3}$, the latter value being compatible with gas densities inferred from the $z\sim 6.3$
GRB afterglow analysis of \cite{gou07}. Neglecting at first entirely the contribution of SNe and HNe
at $z\leq 4$, we plot in Figure~\ref{fig5} the contribution of only these Pop-III HNe
to the neutrino and gamma-ray flux received form the redshift range $4\leq z\leq10$, using the $\alpha=-1$
case as an example.

From this figure \ref{fig5} it is seen that such Pop-III HNe would likely make only a minor contribution to
the diffuse neutrino flux.
However, if the explosion energy is larger, ${\mathcal E}_{\rm POP-III}\sim{10}^{53}$~{\rm erg},
these POP-III HNe by themselves may be able to fit the neutrino data well, without violating the
Fermi-LAT 14\% residual background.  Actually, there is still a larger parameter space for this fitting,
as evidenced by the upper bound shown in Figure~\ref{fig5}, making Pop-III stars more interesting sources
for the present purposes.  Alternatively, our results suggest that their explosion energy is limited to
${\mathcal E}_{\rm POP-III}\lesssim7\times10^{53}$~erg.
%

\begin{figure}
\plotone{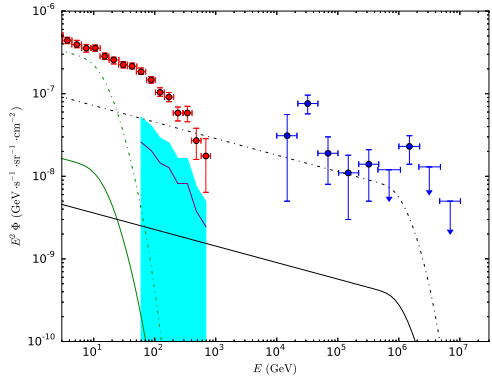}
\caption{Same as Figure~\ref{fig5} but with a CR spectrum index $p=2.2$. For the solid lines the
efficiency $\eta=0.1$ and $\mathcal{E}_{\rm kin}=10^{52.5}$ erg, while for the
dotted-dashed lines, $\eta \mathcal{E}_{\rm kin}$ (i.e. the flux) is multiplied by a factor 20.
\label{fig6}}
\end{figure}

Based on this Pop-III model, we also explore the influence that the initial CR spectral index $p$ has
on the results. Figure~\ref{fig6} shows the case $p=2.2$. Solid lines are for $\eta=0.1$ and dashed lines
are for an enhancement by a factor of 20, while other parameters remain unchanged from Figure~\ref{fig5}.
We see that at high energies the upper bound is lower than that of the $p=2$ case. This means that for $p=2.2$,
under the same constraint given by Fermi-LAT, less neutrino flux can be attributed to the Pop-III stars.
From this point of view, the smaller $p$ are more favorable. We can give a rough constraint of $p\leq2.2$.

\begin{figure}
\plotone{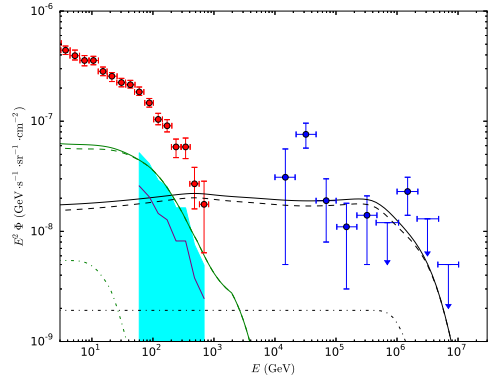}
\caption{An example for two component (low and high redshift) contribution. Black and green solid lines represent
the total diffuse neutrino flux and gamma-ray flux, while the dashed lines are the $z\leq 4$ SNe/HNe and the dotted
lines are the Pop. III SNe. The CR contribution of the Pop. III is instrumental in making this fit more complete
and reasonable, with a fiducial CR efficiency $\eta=0.1$ for both populations.
\label{fig7}}
\end{figure}

Of course there is no reason to assume that only Pop-III HNRs contribute neutrinos in the IceCube energy
range, so a more realistic scenario would involve at least two components contributing to the neutrino
flux, one being the standard SNe and HNe in the usual redshift range $z\leq4$, and the other being the
Pop-III HNe in the $4\leq z\leq10$ range. The occurrence rates in the lower redshift range are observationally
better constrained than for the Pop-III component, but the ratio between the two has large uncertainties.
Since $E_{\nu}^2\Phi_{\nu_i}\propto \eta{\mathcal E}\mathcal{R}$, the ratio between
the Pop-III and Pop-I/II CR contribution can be written as
\begin{eqnarray}
\frac{(E_{\nu}^2\Phi_{\nu_i})_{\rm III}}{(E_{\nu}^2\Phi_{\nu_i})_{\rm I/II}}
&=&\frac{\eta{\mathcal E}_{\rm HN}\mathcal{R}_{\rm III}(\bar{z}_{\rm III})}{\eta{\mathcal E}_{\rm SN}0.97\mathcal{R}_{\rm I/II}(\bar{z}_{\rm I/II})+\eta{\mathcal E}_{\rm HN}0.03\mathcal{R}_{\rm I/II}(\bar{z}_{\rm I/II})}\nonumber\\
&\times&\frac{(1+\bar{z}_{\rm I/II})^2\sqrt{\Omega_M(1+\bar{z}_{\rm I/II})^3+\Omega_\Lambda}}{(1+\bar{z}_{\rm III})^2\sqrt{\Omega_M(1+\bar{z}_{\rm III})^3+\Omega_\Lambda}}
\frac{f_{pp}^{\rm III}}{f_{pp}^{\rm I/II}}.
\label{eq:cr13}
\end{eqnarray}
Here, as seen also from Fig.~8, the first term is $\sim0.3$, where the typical redshifts of POP I/II and POP III
remnants are $\bar{z}_{\rm I/II}\sim1$ and $\bar{z}_{\rm III}\sim5$, respectively.
The second term gives $\sim0.02$ and the last term is $\sim20$, leading to the ratio is $\sim0.1$.
The star formation rate will be lower for Pop-III than it is at low redshifts.  On the other hand,
$\eta{\mathcal E}\mathcal{R}(z)$ could be substantially larger due to a plausibly larger kinetic energy.
Thus, introducing this Pop-III contribution potentially allows one to produce a fraction approaching
unity of the low and high redshift IceCube neutrino flux, without violating the Fermi residual
gamma-ray background.  An example with these two components is shown in Figure~\ref{fig7}, where
the Pop-III/Pop-I/II efficiency is set to the same $\eta=0.1$. We can see that Pop-III HNe can contribute
an interesting fraction to the diffuse neutrino flux, providing a good argument in favor of SBG scenario.

\begin{figure}
\plotone{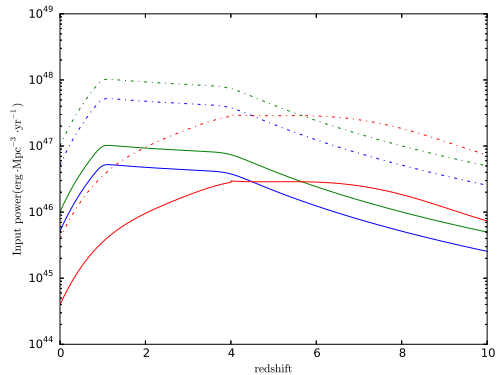}
\caption{The redshift evolution of the CR input power $\eta{\mathcal E}\mathcal{R}(z)$ for SNe (blue solid),
HNe (green solid) and Pop-III HNe (red solid), assuming $\eta=0.1$. Dashed lines are the total power
$\epsilon_pQ_{\epsilon_p}$.
\label{fig8}}
\end{figure}

Figure~\ref{fig8} shows the CR input power $\epsilon_pQ_{\epsilon_p}$ versus redshift for this fit,
the Pop-III being subdominant (red solid line). However, with a higher density $\bar{n}$, an $f_{pp}=1$
(at the low energy end) is obtained for Pop-III, while $f_{pp}\sim0.3$ for SBGs and $f_{pp}\sim0.03$ for SFGs,
which compensates for the lower star formation rates of POP-III stars.  (Note that the assumed energy injection
might be optimistic because of the uncertainty in $f_{pp}$ and $\xi_{\rm SBG}$.)

\section{Discussion and Conclusions}

In this work we have considered the possible contribution of standard core collapse supernovae and
hypernovae to the observed IceCube diffuse neutrino background, including two aspects which, to
our knowledge, have not been previously considered in the literature. One of these aspects is
the time history of the SNR evolution, which affects the effective total cosmic ray spectral output.
Considering the contribution of SNRs and HNRs in the redshift range $z\leq 4$ only, and considering
a CR spectral injection power law slope $p=2$ and a CR injection efficiency which during the
Sedov-Taylor phase tapers off as a time power law $\propto t^\alpha$ with $\alpha \leq 0$,
we performed detailed calculations of the resulting diffuse neutrino flux and the diffuse
gamma-ray flux contributed by SNRs of both the above types.
It remains difficult to account for the neutrino data point at 30 TeV as well as for the rest
of the higher energy points with any single power law injection spectrum,
as pointed out by several authors \citep{mur16, bec15, kis15}.
However, if we do not include the 30 TeV data point, e.g. by implicitly assuming that it is an extra
component, perhaps due to hidden sources \citep{mur16}, we find that the neutrino flux contributed
by SNe and HNe in SFGs at $z\leq 4$ can explain about 50\% of the remaining IceCube observations,
without violating the fraction of 14\% of the diffuse gamma-ray flux which, according to the Fermi-LAT
collaboration, cannot be attributed to unresolved AGNs of known types \citep{ack12, dim16}.

In view of the observations showing that core-collapse GRBs are associated with SN/HN type Ib/c \citep[e.g.][]{Gehrels+09araa}, it is pertinent to inquire what may be the potential contribution
of such GRBs to the diffuse neutrino and gamma-ray backgrounds discussed here. Although so far there 
only about seven such associations verified through their spectrum, and a few tens verified through light 
curves \citep[e.g.][]{can16}, which is a small fraction of all GRBs ($\ll 10\%$), the association is believed to apply to all
long GRBs. Note that the SNe associated with GRBs are mostly hypernovae, and all are of type Ib/c, 
while here in our work the SNe and HNe considered are of all the core collapse types II+Ib/c.
At low redshifts, HNe are a fraction $5\%-10\%$ of all core collapse SNe, and only a fraction of 
order $5\%-10\%$ appears associated with GRBs \citep{mod15, arc10, sod10}.  Since the average GRB jet 
beaming correction is roughly 1/100 \citep{gue05}, the fraction of detected GRBs which could be positionally 
associated with the neutrinos or gamma-rays discussed here is $\lesssim (1/10)(1/10)(1/100)\sim 10^{-4}$, 
which is negligible in our scenario.

Since at best only a fraction of the IceCube neutrinos can be explained by SNe and HNe in the
redshift range $z\leq 4$, the other new aspect considered here is the possible contribution of
Pop-III HNe at $4\leq z \leq 10$.
\citet{mur16} discussed a possible contribution of very-high-redshift injections.
\citet{cha16} recently argued that the IceCube generic neutrino sources may reside at high redshifts,
but their focus is not on Pop-III remnants.
In our case we specifically considered Pop-III HNRs as main sources for both the low and high redshift
neutrinos and gamma-rays, and we have used a specific model of the EBL evolution in the Pop-III redshift
range \citep{ino13}. The Pop-III SNRs are expected to be more energetic than their lower redshift
counterparts, and could contribute significantly to the total CR output.
The neutrinos will reach the observer unimpeded, but the gamma-rays produced at high redshifts need
to travel longer and have a greater chance to be attenuated by interacting with the EBL.

We find that, aside from the 30 TeV data point, a neutrino flux compatible with IceCube could be 
obtained by considering a substantial contribution from a Pop-III component, which does not violate 
the residual diffuse gamma-ray background constraint. Figure~\ref{fig7} is an example of such a fit, where the fiducial parameters ${\mathcal E}_{\rm POP-III}=10^{52.5}$ erg and $\eta=0.1$ 
are used for Pop-III HNRs, which here contribute $\sim 10\%$ of the total neutrino flux.
A larger neutrino contribution which still respects the Fermi bounds, as shown in Fig.5,
would require ${\mathcal E}_{\rm POP-III}\sim{\rm a~few} \times10^{53}$ ~erg for the Pop-III rate 
model of \citet{ino13}, or larger for more pessimistic rate models such those of \citet{tre09, wis12}.
Thus we conclude that it will be difficult for Pop-III HNRs to dominantly contribute to the observed 
high-energy backgrounds unless the explosion energy is ${\mathcal E}_{\rm Pop-III}\gtrsim 10^{53}$~erg.
On the other hand, the explosion energy for supermassive stars at very-high-redshifts could be larger 
than $\sim{10}^{55}$~erg \citep{mat15}.   
Note that the relative ratio of the neutrino fluxes contributed by the high and low redshift components
are largely uncertain.  If the fraction of Pop-III HNRs were even higher than assumed here, there would 
be even less arriving high-energy gamma-rays, further weakening the tension with Fermi-LAT.

\acknowledgements
We thank Nick Senno for discussions, and we acknowledge support by the National Basic Research Program
of China (973 Program grant 2014CB845800 and the National Natural Science Foundation of China grant 11573014
(D.X. and D.Z.G.), by the program for studying abroad supported by China Scholarship Council (D.X.),
by Pennsylvania State University (K.M.) and by NASA NNX13AH50G (P.M.).

\par

\clearpage

\end{document}